\documentclass[%
 reprint,
 amsmath,amssymb,
 aps,
pra,
]{revtex4-1}

\usepackage{graphicx}
\usepackage{dcolumn}
\usepackage{bm}

\begin{document}
\preprint{APS/123-QED}
\title{Two-way single-photon-level frequency conversion between 852nm and 1560nm for connecting cesium D2 line with the telecom C-band}
\author{Kong ZHANG${}^{1}$}
\author{Jun HE${}^{1,2}$}
\author{Junmin WANG${}^{1,2,}$}
\email{Corresponding Author: wwjjmm@sxu.edu.cn}
\affiliation{{${}^{1}$State Key Laboratory of Quantum Optics and Quantum Optics Devices, and Institute of Opto-Electronics, \\Shanxi University, Tai Yuan 030006, People's Republic of China}\\{${}^{2}$Collaborative Innovation Center of Extreme Optics of the Education Ministry and Shanxi Province, Shanxi University, Tai Yuan 030006, People's Republic of China}}
\begin{abstract}
A compact setup for two-way single-photon-level frequency conversion between 852 nm and 1560 nm has been implemented with the same periodically-poled magnesium-oxide-doped lithium niobate (PPMgO:LN) bulk crystals for connecting cesium D2 line (852 nm) to telecom C-band. By single-pass mixing a strong continuous-wave pump laser at 1878 nm and the single-photon-level periodical signal pulses in a 50-mm-long PPMgO:LN bulk crystal, the conversion efficiency of $ \sim $ 1.7 \% ($ \sim $ 1.9 \%) for 852-nm to 1560-nm down-conversion (1560-nm to 852-nm up-conversion) have been achieved. We analyzed noise photons induced by the strong pump laser beam, including the spontaneous Raman scattering (SRS) and the spontaneous parametric down-conversion (SPDC) photons, and the photons generated in the cascaded nonlinear processes. The signal-to-noise ratio (SNR) has been improved remarkably by using the narrow-band filters and changing polarization of the noise photons in the difference frequency generation (DFG) process. With further improvement of the conversion efficiency by employing PPMgO:LN waveguide, instead of bulk crystal, our study may provide the basics for cyclic photon conversion in quantum network.

\end{abstract}

\maketitle


\section{Introduction}
\textit{}
Quantum state transmission is the key to realize quantum network. In 1997, Cirac  \emph{et al} [1] proposed a scheme to utilize photons for ideal quantum transmission between atoms located at spatially separated nodes of a quantum network. Implementation of long-distance quantum information network requires the quantum nodes which store and process quantum information and the low-loss optical fiber networks which connect the nodes by flying photonic qubits. The photonic frequency of quantum nodes are usually much higher than that of flying photonic qubits transmitted in the optical fiber networks. Therefore, it needs quantum frequency conversion (QFC) [2,3] of the photonic frequency between the quantum nodes and the flying photonic qubits. Several qubit platforms, such as trapped ions [4], trapped cold single atom or cold atomic ensemble [5,6], quantum dots [7], and etc have been demonstrated for building quantum nodes for the quantum network. In the experiment, many groups used single photons emitted from qubit platforms convert ranging from 637 and 980 nm to the telecom bands with SNR of $10^2$
order with efficiencies between $ \sim $10-70\% [8-19].

Most of the above-mentioned conversions aimed at red spectral ranges converted to telecom bands. However, inverse conversions are also important. Moreover, some groups have demonstrated the photons' up-conversion experiments [20,21]. In 2018, Wright \emph{et al} [22] reported two-way photonic interface for linking 422 nm to the telecom C-band, and achieved up-conversion (down-conversion) at single-photon level with conversion efficiency of $ \sim $ 9.4\% ($ \sim $ 1.1\%), and SNR of $ \sim $ 39 ($ \sim $ 108). The up-conversion single-photon detector (SPD) using this scheme can extend the well-developed silicon based single-photon detector to telecom C-band.

Now we consider a simple quantum network shown in Fig.1 for single-photon distributing and storing. The left node is single-cesium (Cs)-atom optical tweezer based 852-nm single-photon source [23-25], while the right node is cold or hot Cs atomic ensemble. These two nodes can be connected by ultralow loss 1560-nm telecom single-mode fiber. By employing QFC1 module, the 852-nm single photons emitted by the left node can be converted to 1560-nm single photons, which served as the flying photonic qubits for single-photon distributing via long-distance 1560-nm telecom fiber link. Then by employing QFC2 module, 1560-nm single photons can be converted back to 852-nm single photons, which can interact with the right node via Raman-type memory or the electromagnetically induced transparency (EIT) - type memory protocols for single-photon storing.

In this paper, we demonstrated two-way down-conversion and up-conversion between 852 nm and 1560 nm based on single-pass periodically-poled magnesium-oxide-doped lithium niobate (PPMgO:LN) bulk crystal for connecting 852-nm Cs D2 transition and 1560-nm telecom C-band at single-photon level (one mean photon per pulse). The efficiency is mainly limited by spatial mismatch in the bulk crystal, and it is hard to effectively use whole bulk crystal because the beams are focused. These problems can be avoided by using a waveguide, and the efficiency will be improved. So we pay more attention to improving SNR. First, we analyzed the noise photons induced by a strong pump laser beam in PPMgO:LN crystal, such as the spontaneous Raman scattering (SRS) and the spontaneous parametric down-conversion (SPDC) photons, as well as the photons generated in the cascaded nonlinear processes. Then, we enhance SNR by using the narrow band-pass filters. Moreover, by rotating the polarization of the output photons, the noise photons are filtered by PBS at the expense of extraction efficiency in difference frequency generation (DFG) experiment.

\begin{figure}[h]
\vspace{-0.17in}
\centerline{
\includegraphics[width = 90mm]{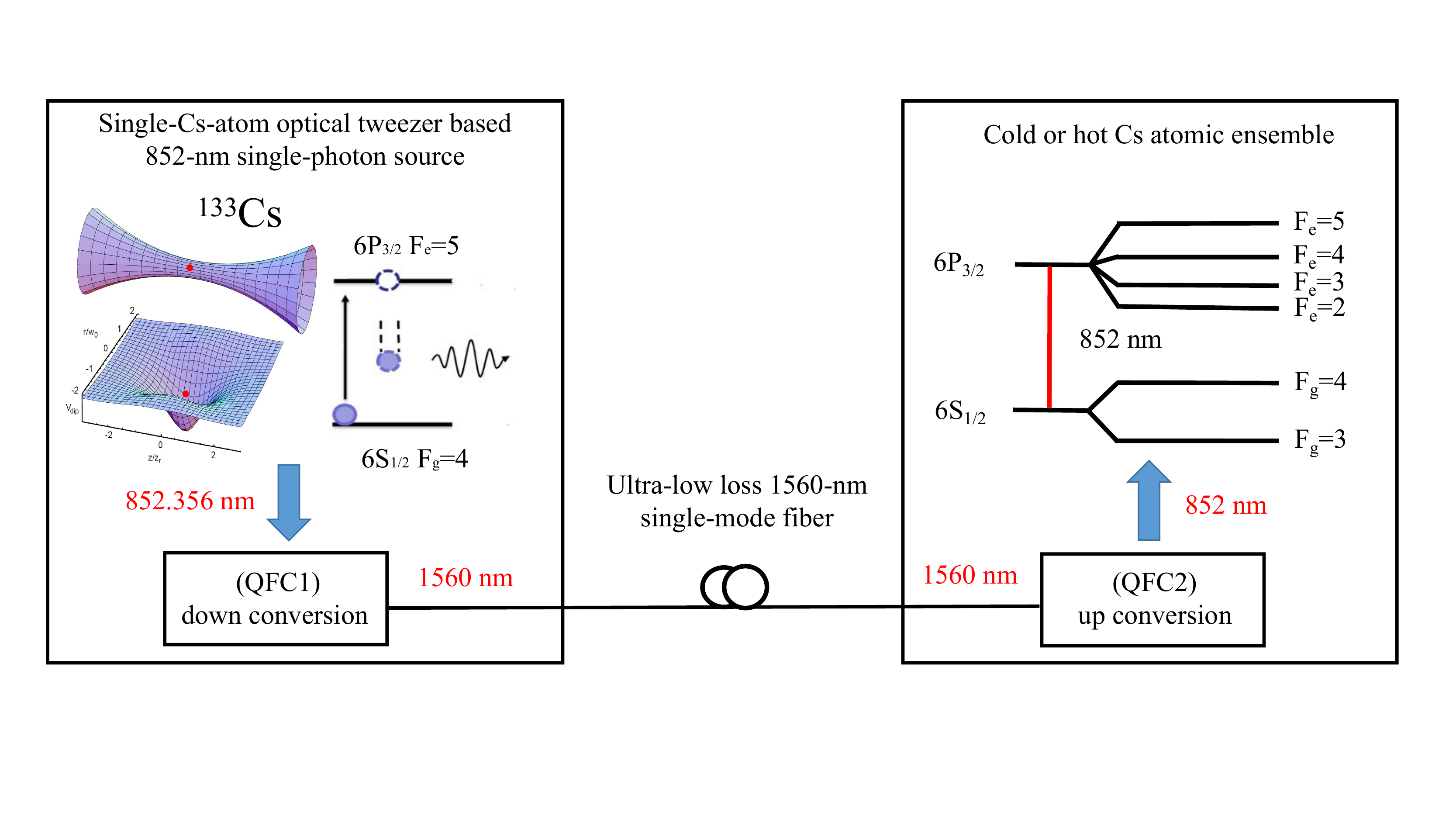}
} \vspace{-0.27in} 
\caption{Schematic diagram of two-way QFC for distributing and storing single photons. The left node is single-Cs-atom optical tweezer based 852-nm single-photon source, while the right node is cold or hot Cs atomic ensemble. These two nodes can be connected by low-loss 1560-nm telecom single-mode fiber. By employing QFC1 module the 852-nm single photons emitted by the left node can be converted to 1560-nm single photons served as the flying photonic qubits, and then can be converted by employing QFC2 module back to 852-nm single photons which interact with the right node.}
\label{Fig 1}
\vspace{-0.1in}
\end{figure}

\section{DFG from 852-nm to 1560-nm photons: Experiment and results}

In this part, we convert an attenuated coherent state from 852 nm to 1560 nm by DFG and enhance the SNR by using the narrow band-pass filters and changing the polarization of the noise photons. At the same time, we also characterize the noise photons induced by the strong pump beam. Fig.2 (a) is diagram of single-photon-level down-conversion. The experimental setup is shown in Fig.2 (b). A master-oscillator power amplifier (MOPA) consists of a compact distributed feedback (DFB) diode laser and a Thulium-doped fiber amplifier (TmDFA) (PreciLasers Corp, Shanghai, China) can provide watt-level single-frequency narrow-band 1878-nm pump light. As we know, the wavelength of commercial TmDFA is 1920 nm - 2200 nm, while the 1878-nm TmDFA used in experiment is produced with special technique. An optical isolator is used to restrain the laser feedback, thus ensuring stable operation of TmDFA. To simulate single-photon pulses, a compact DFB diode laser at 852 nm, an acoustic-optical modulator (AOM) and a strong attenuator are employed to chop and attenuate the 852-nm continuous-wave laser beam to 1-MHz repetition rate and 500-ns duration square-wave periodical optical pulses with mean photon number per pulse is $ \sim $ 1. Then, a 50-mm-long PPMgO:LN bulk crystal (the thickness: 0.5 mm; the poling period: 23.4 ${{\mu}}$m; Type-0 quasi-phase matching; The both flat ends of the crystal were anti-reflection coated for 852 nm, 1560 nm, and 1878 nm, and the residual reflectivity R $<$ 0.2\%) fabricated by HC Photonics is employed for single-photon level frequency conversion. The crystal is placed in a home-made oven, which is made of red copper and precisely temperature stabilized by using a temperature controller (Newport Corp., Model 350B, the temperature stability is $ \sim $ $\pm$ 0.1$^{\circ}$C). We can achieve the optimized phase matching by adjusting the temperature of the crystal.

\begin{figure}[h]
\setlength{\belowcaptionskip}{-0.5cm}
\vspace{-0.1 in}
\centerline{
\includegraphics[width=75mm]{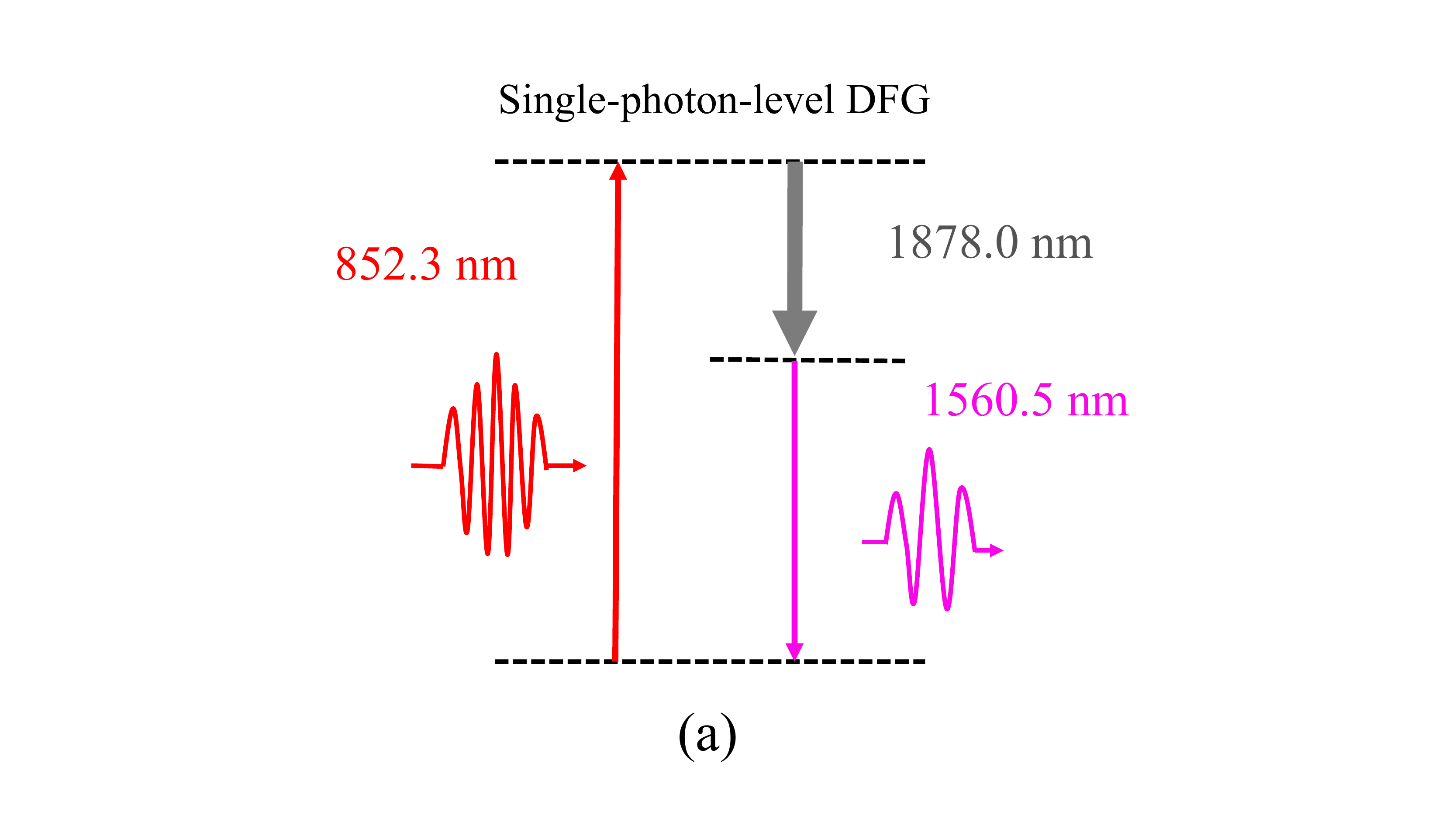}
} \vspace{-0.3in} 

\vspace{-0.1in}
\end{figure}

\begin{figure}[h]
\setlength{\belowcaptionskip}{-0.5cm}
\vspace{0 in}
\centerline{
\includegraphics[width=95mm]{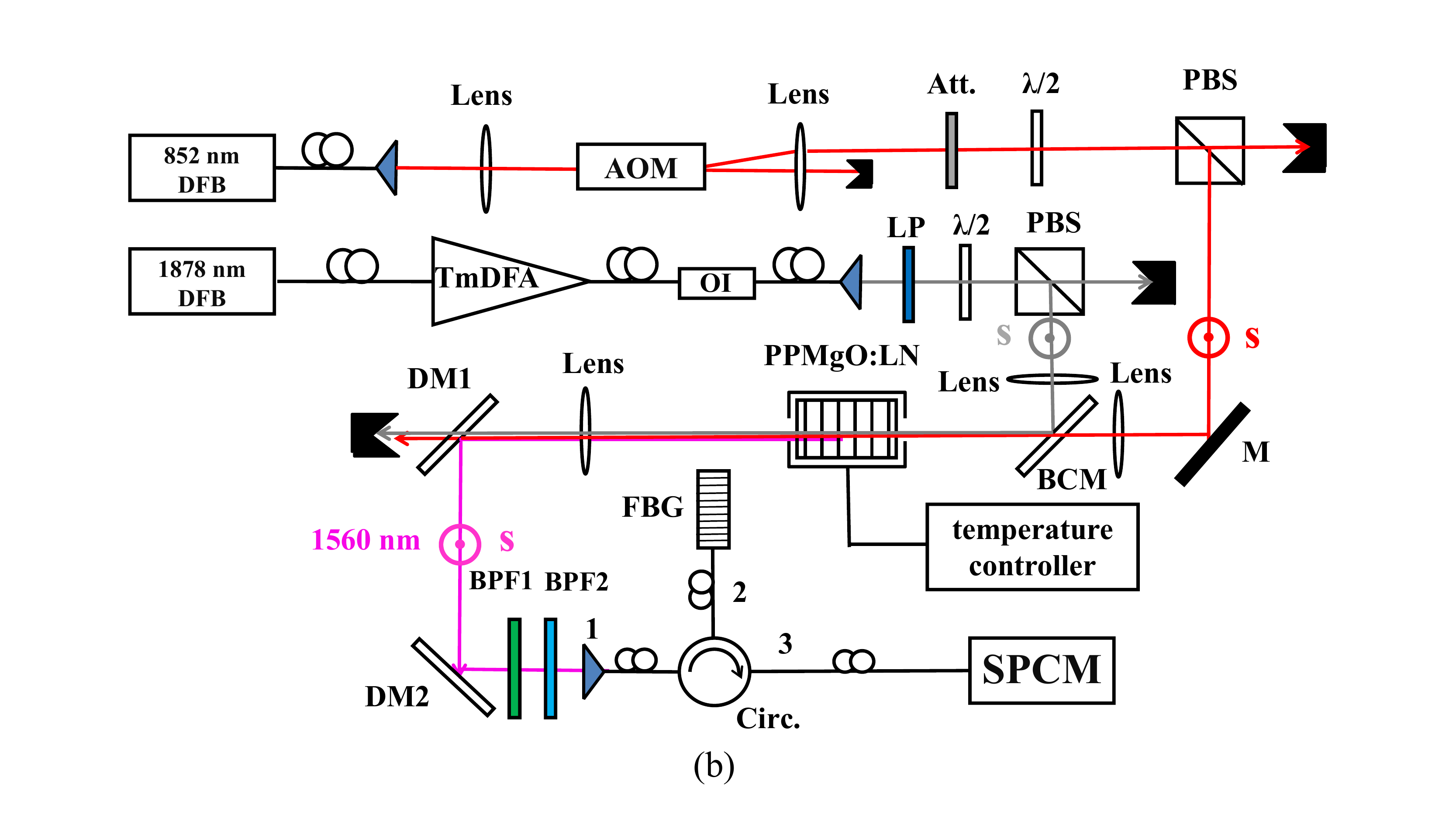}
} \vspace{0in} 
\caption{(a) Diagram of single-photon-level down-conversion. (b) The experimental setup for DFG. The s-polarized 852-nm photons are converted to 1560-nm photons with PPMgO:LN crystal by DFG pumped by the s-polarized 1878-nm strong laser beam. DFB: distributed feedback diode laser; TmDFA: Thulium-doped fiber amplifier; OI: optical isolator; AOM: acousto-optic modulator; ${{{\lambda}/2}}$: half-wave plate; PBS: polarization beam splitter cube; Att.: attenuator; s: s polarization; BCM: beam-combination mirror; DMs: 1560-nm high-reflection and 852-nm and 1878-nm high-transmission dichromatic mirrors; BPFs: 1560-nm band-pass filters; Circ.: fiber circulator; FBG: fiber Bragg grating filter ($ \sim $ 0.3 nm); SPCM: single-photon counting module; LP: low pass filter.}
\label{Fig 2}
\vspace{0.3in}\end{figure}

First, we find the optimized phase matching condition by adjusting the PPMgO:LN crystal's temperature in single-pass DFG experiment. When the optical power of the 1878-nm pump beam was $ \sim $ 450 mW, the phase matching is achieved at a temperature of $ \sim $ 78.9$^{\circ}$C, as shown in Fig. 3.

\begin{figure}[h]
\setlength{\belowcaptionskip}{-0.5cm}
\vspace{-0.05 in}
\centerline{
\includegraphics[width=100mm]{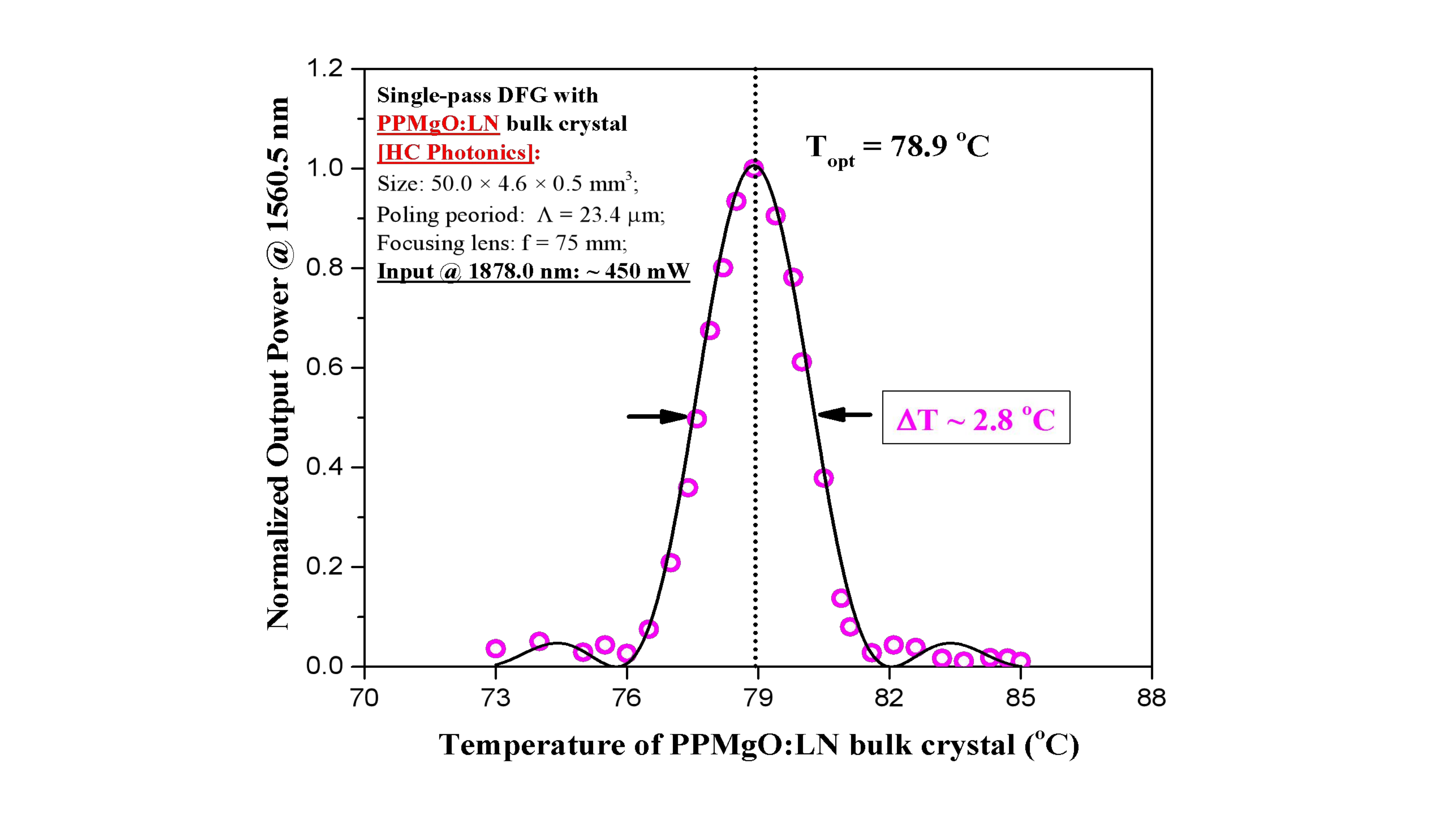}
} \vspace{-0in} 
\caption{Dependence of 1560.5-nm output power upon the crystal's temperature in the case of DFG. The open circles represent the experimental data, while the solid line is theoretical fitting curve. The optimal quasi-phase-matching temperature is $ \sim $ 78.9 $^{\circ}$C, and the temperature bandwidth is $ \sim $ 2.8 $^{\circ}$C.}
\label{Fig 2}
\vspace{0.2in}
\end{figure}

\begin{figure}[h]
\vspace{-0.15in}
\centerline{
\includegraphics[width=95mm]{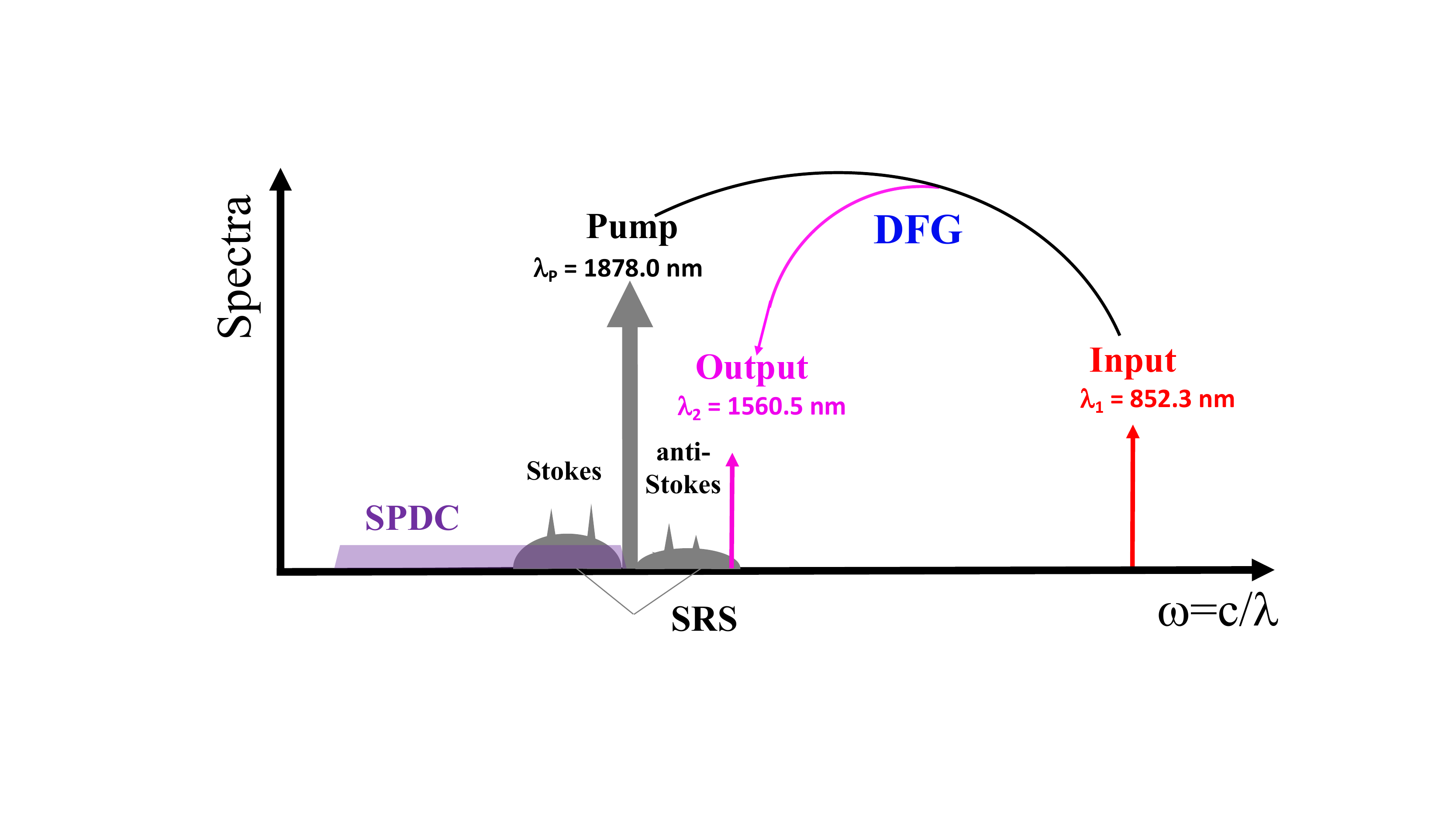}
} \vspace{-0.35in} 
\caption{Overview of DFG photonic conversion process from 852.3 nm to 1560.5 nm with the 1878.0-nm pump laser and the strong pump induced nonlinear processes. SPDC: the spontaneous parametric down-conversion (marked by the purple part); SRS: the spontaneous Raman scattering (Stokes photons and anti-Stokes photons, and marked by the dark grey part). The narrow BPFs and the ultra-narrow FBG filter can be utilized to remove the noise photons located around the target output at 1560.5 nm.}
\label{Fig 4}
\vspace{-0.1in}
\end{figure}

Then, we pay more attentions to the noise photons induced by the 1878-nm strong pump beam. In the experiment of single-photon-level frequency conversion, pump laser with long wavelength can minimize the influence of noises [15, 16]. Fig. 4 shows the overview of DFG photonic conversion process from $ {\lambda}_{1}$= 852.3 nm to $ {\lambda}_{2}$= 1560.5 nm with the $ {\lambda}_{p}$= 1878.0 nm pump laser and the strong pump induced nonlinear processes (SPDC and SRS). First, the wavelength of SPDC photons (marked by the purple part) are longer than the pump laser, it is far away from the targeted photons, so we assume that SPDC photons can be excluded in our case. Then SRS photons (marked by the dark grey part) distribute on both sides of the pump light, and the target photons locate in weak anti-Stokes region, so the main noise photon source in this process are anti-Stokes photons. The narrow BPFs and the ultra-narrow FBG can be utilized to remove the noise photons located around the target output. In addition, the TmDFA with a low gain at 1878 nm in experiment is not a commercial one, so it may generate more noises with higher pumping diode laser's power, which will reduce the SNR.

In the experiment, the conversion efficiency be defined as  $\eta$=${{N}_{out}}$/${{N}_{in}}$, we define ${{N}_{in}}$ as the count rate of the input signal photons in the front the crystal, and ${{N}_{out}}$ as the count rate of the output target photons at the end of the crystal. The influence of noise photons at target wavelength induced by the strong pump light is inevitable. Therefore, we define ${{N}_{out}}$=${{N}_{p+s}}$-${{N}_{p}}$, where ${{N}_{p+s}}$ is the count rate of the detector at 1560 nm when both pump and signal photons coupled into the crystal, and ${{N}_{p}}$ is the count rate of the detector at 1560 nm when only pump light is present in the crystal (signal photons are blocked). Zaske \emph{et al} [15] defined the conversion efficiency in the same way. Fig. 5 shows the DFG conversion efficiency and SNR with different filters versus pump power. The SNR can be slightly improved by using narrower filters. Fig. 5 (a) is the result for a BPF with $ \sim $ 50-nm FWHM bandwidth (T $ \sim $92\%@1560 nm and OD=4 for 852 nm), and the SNR is 19.5. Then the SNR can be improved to 31.3 by using a BPF with $ \sim $ 12-nm FWHM bandwidth (T  $ \sim $ 74\%@1560 nm, OD=4 for 1878 nm and OD=5 for 852 nm) combined a FBG filter with $ \sim $ 0.3-nm FWHM bandwidth (R $ \sim $92\% @1560 nm) in Fig. 5 (c). Therefore, the SNR will be increased by using narrower bandwidth of the filters theoretically. However, when using a very narrow filter (the bandwidth is much smaller than the filter used in our experiment), the SNR is not significantly improved. The effect of temperature on noises is crucial [28]. Later we will choose a new poling period to make the quasi-phase matching temperature of the crystal much lower than that of now, therefore we can expect that we should get much lower anti-Stokes (Raman scattering) noise photons counting. In the result of conversion efficiency, the solid lines represent the fitting to Eq. (1):

\vspace{-0.2 in}
\hspace*{-0.65cm} 

\begin{eqnarray}
{\eta({P_{p}})}={\eta_{max}}{sin^{2}}(L\sqrt{{\eta_{n}}{P_{p}}})
\end{eqnarray}

Where $L$ is the length of the non-linear crystal, $ P_{p}$ is the injected pump power, and $ n$ is a conversion parameter characteristic to the specific device. However, the curvature of curve (a) and (b) are different from (c), which may be due to the bandwidth of filters, and the efficiency in Fig. 5 (c) is close to saturation, which is mainly caused by the photons consumption ratio of signal photons to pump photons. The maximum DFG internal conversion efficiency is $ \sim $1.7 \% when signal level is set to $ \sim $ ${{10}^{6}}$ photons ${{s}^{-1}}$ with a mean photon number per pulse is $ \sim $ 1. The efficiency is mainly limited by spatial mismatch of the signal and the pump beams in the crystal [29]. The waist spots radii and the divergence angles of 852 nm and 1878 nm are different. Moreover, we cannot use the whole crystal length because the beams are focused. These two problems can be effectively avoided by using a waveguide, and the efficiency will be improved. Last, the inverse sum-frequency generation (SFG) from 1560 nm to 852 nm photons will be occurred under the similar condition. Therefore, we think it is inevitable that the SFG will be happened in the DFG experiment.  This
 \begin{figure}

\centering

{

\begin{minipage}[h]{0.5\textwidth}

\includegraphics[width=95mm]{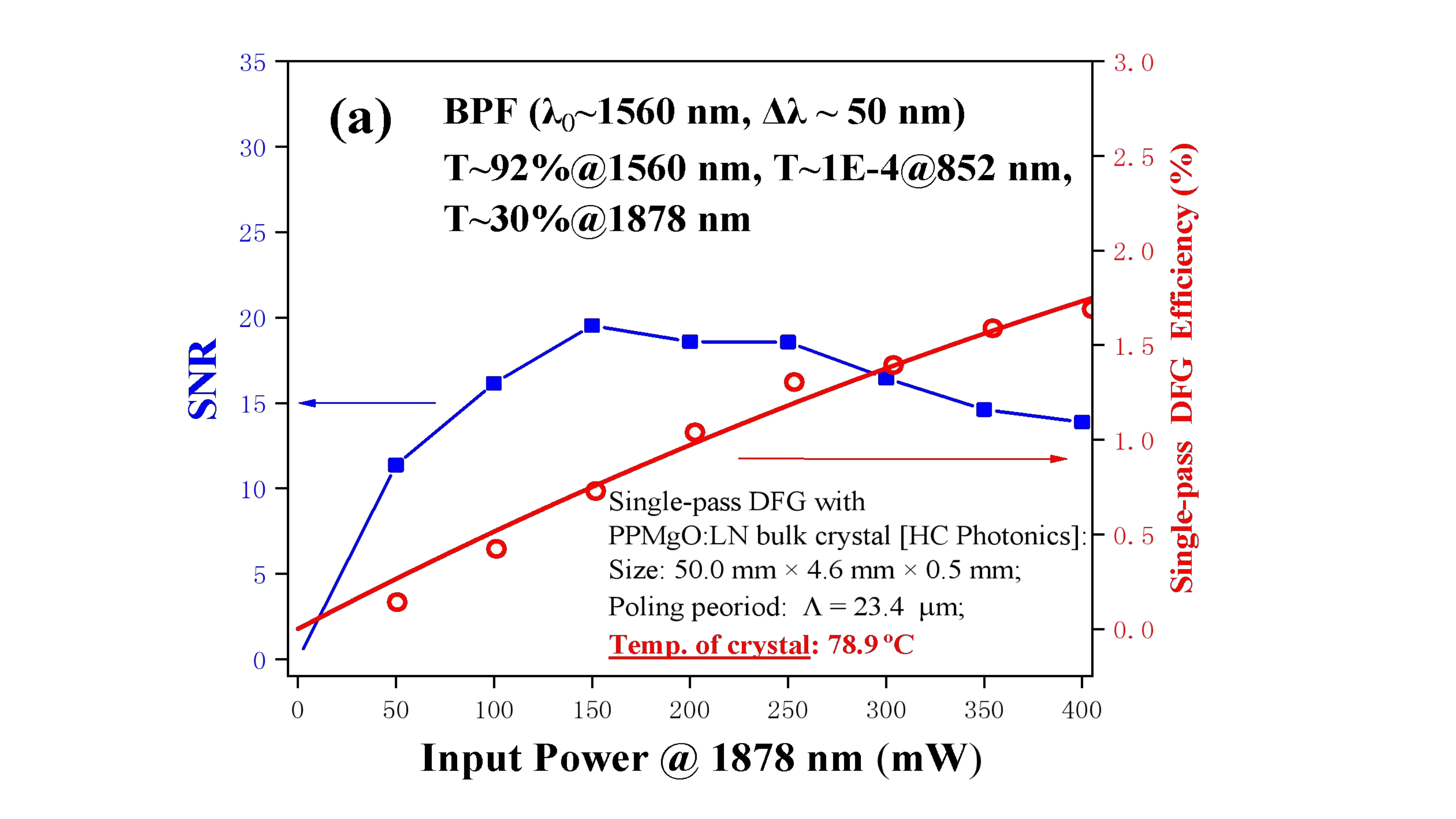}
\vspace{-0.05 in}

\end{minipage}

}

{

\begin{minipage}[h]{0.5\textwidth}

\includegraphics[width=95mm]{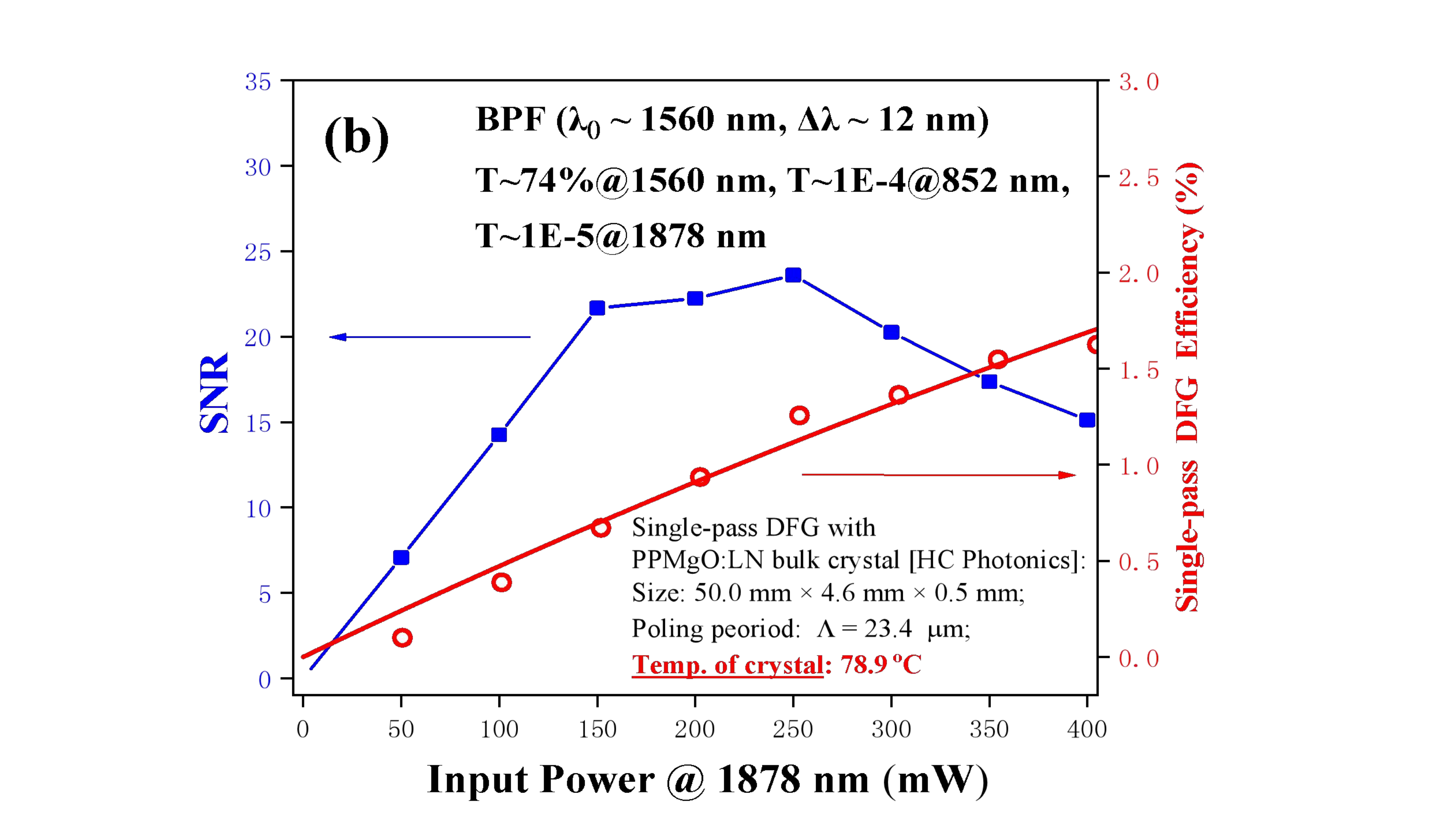}

\end{minipage}
\vspace{0.05 in}

}
{

\begin{minipage}[h]{0.5\textwidth}

\includegraphics[width=95mm]{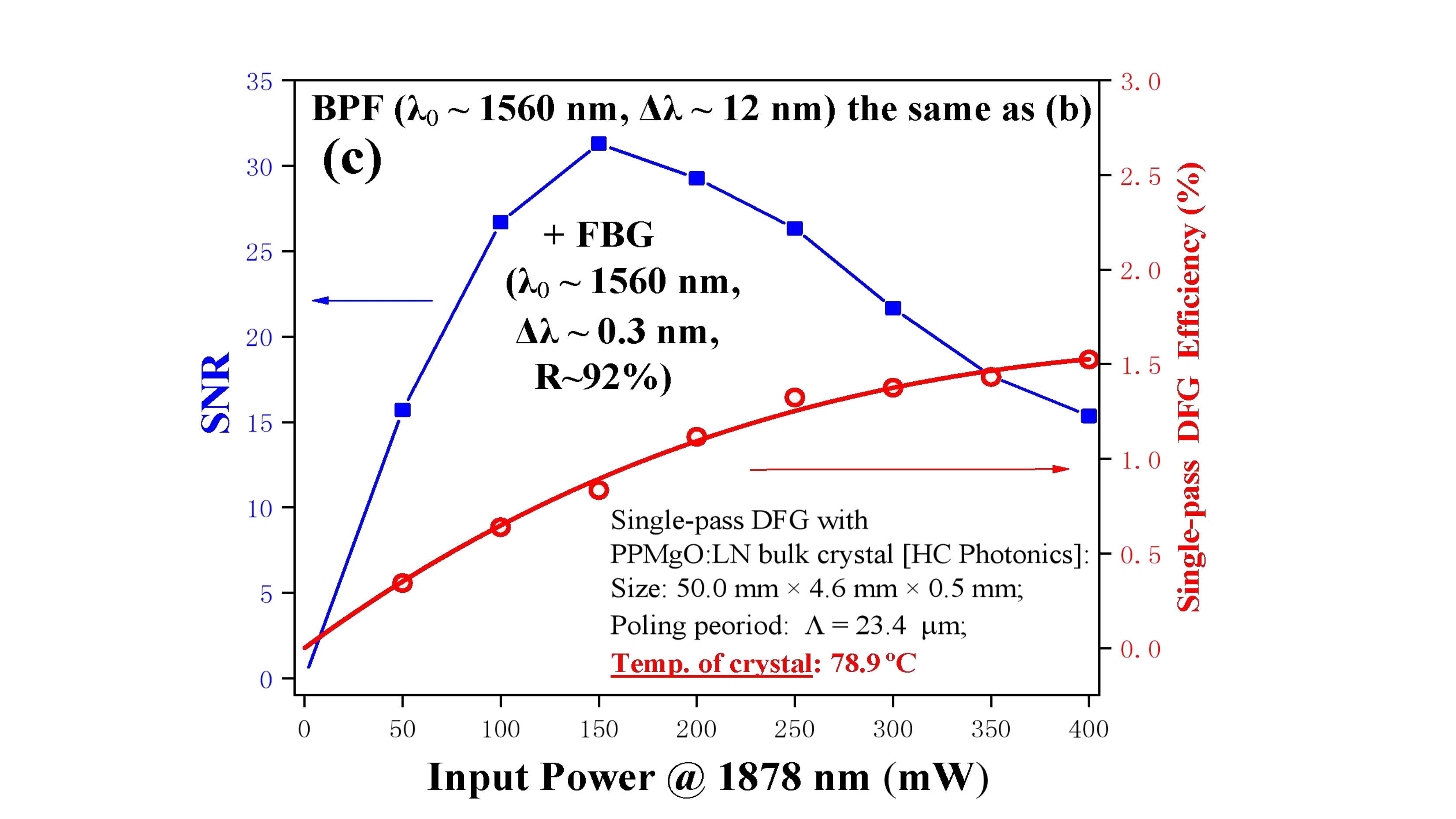}
\vspace{-0.05 in}
\end{minipage}

}

 \caption{DFG conversion efficiency and SNR with different filters versus the 1878-nm pump power. The blue solid squares represent the SNR, and the red open circles represent the DFG conversion efficiency. (a) The results for a BPF with $ \sim $50-nm FWHM bandwidth; (b) The results for a BPF with $ \sim $12-nm FWHM bandwidth; (c) The results for combination of a BPF with $ \sim $12-nm FWHM bandwidth and a FBG filter with $ \sim $0.3-nm of a FWHM bandwidth.DFG conversion efficiency and SNR with different filters versus the 1878-nm pump power. The squares represent the SNR, and the circles represent the DFG conversion efficiency. (a) The results for a BPF with $ \sim $ 50-nm HMFW bandwidth; (b) The results for a BPF with $ \sim $ 12-nm HMFW bandwidth; (c) The results for combination of a BPF with $ \sim $ 12-nm HMFW bandwidth and a ultra-narrow-band FBG filter with $ \sim $ 0.3-nm HMFW bandwidth.}
\vspace{-0.15 in}
\end{figure}

\vspace{-0.15 in}

 \noindent means that some of the 1560 nm photons generated by the DFG may be further converted back to 852 nm photons in the crystal. the inverse conversion efficiency is much less than the target conversion efficiency, that factor may also limits our DFG efficiency.

Although the PPMgO:LN crystal we used is type-0 quasi-phase matching (852-nm, 1878-nm and 1560-nm photons are all with s polization), the noise photons (mainly are anti-Stokes photons of SRS induced by the strong 1878-nm pump laser beam) may be elliptically polarized with a special preponderant direction due to inhomogeneity of the crystal. Based on this point, we try to improve SNR by changing the polarization of output photons by using of the setup shown in Fig. 6 (a). The output target 1560-nm photons and some anti-Stokes photons together pass through all the DMs and BPF as well as FBG filter, then we change their polarization by ${{{\lambda}/2}}$ plate and count the reflected s-polarization photons of PBS. The Circ and FBG filter are both single-mode modules, and are not sensitive to polarization. Fig. 6 (b) shows SNR versus target 1560-nm photons' extraction efficiency. By rotating ${{{\lambda}/2}}$ plate, the photon counts of reflection channel are reduced, and the preponderant direction of noise photons are changing to p. SNR can be improved to 58.3 with $ \sim $ 70\% of extraction efficiency. We can derive the polarization rotation angle according to different extraction efficiency by Malus law. When the extraction efficiency is $ \sim $ 70\%, the polarization rotation angle is $ \sim $ 33$^{\circ}$.

Fig. 6 (c) shows the results for SNR versus 1878-nm pump power. The blue solid squares represent SNR without the ${{{\lambda}/2}}$ plate and PBS, and the red open circles represent SNR with the ${{{\lambda}/2}}$ plate and PBS at the optimized polarization direction (extraction efficiency  $ \sim $ 70\%). The

\begin{figure}[h]
\setlength{\belowcaptionskip}{-0.5cm}
\vspace{-0.1 in}
\centerline{
\includegraphics[width=80mm]{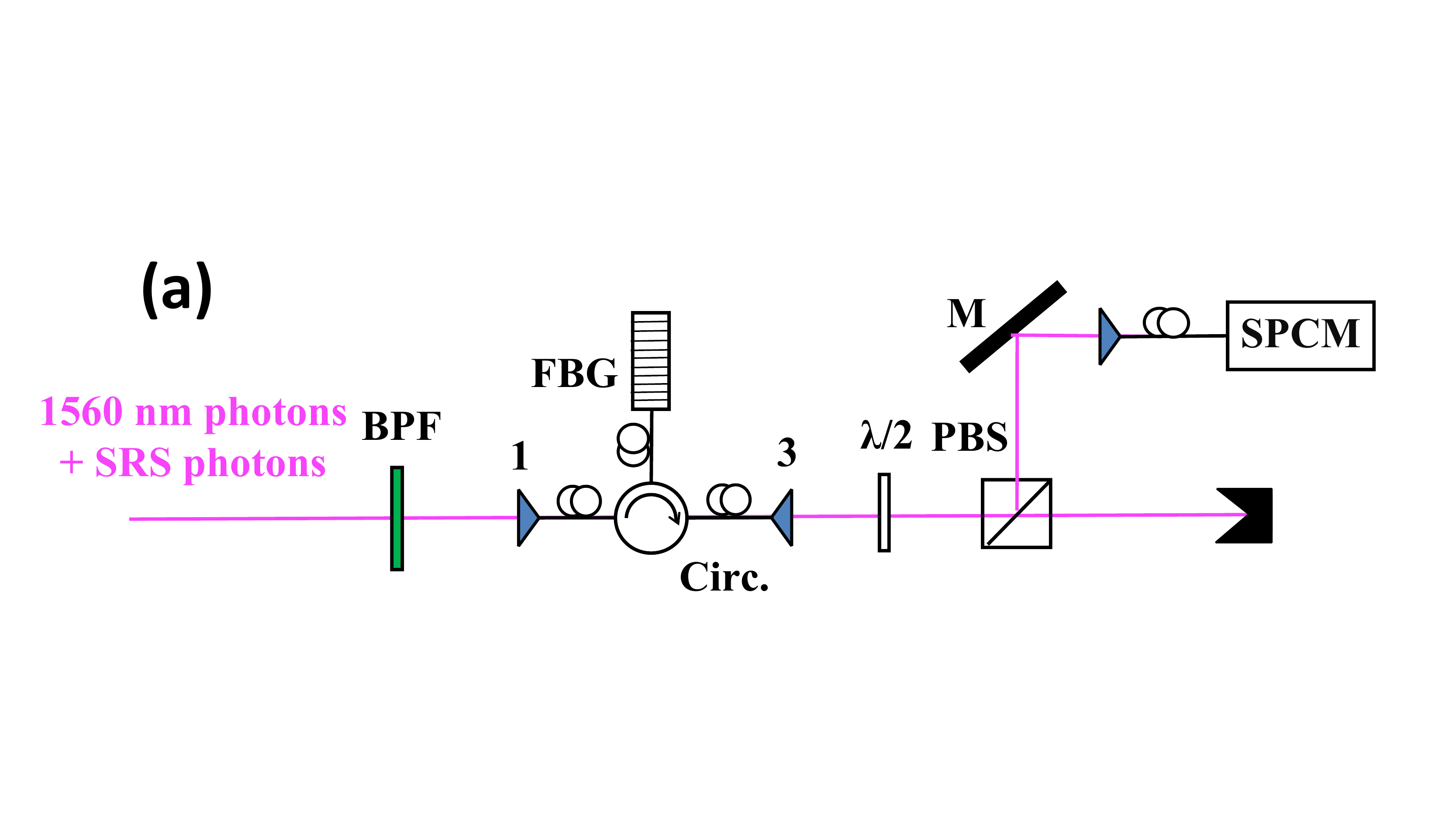}
} \vspace{-0.1in} 

\end{figure}

\begin{figure}[h]
\setlength{\belowcaptionskip}{-0.5cm}
\vspace{-0.5 in}
\centerline{
\includegraphics[width=95mm]{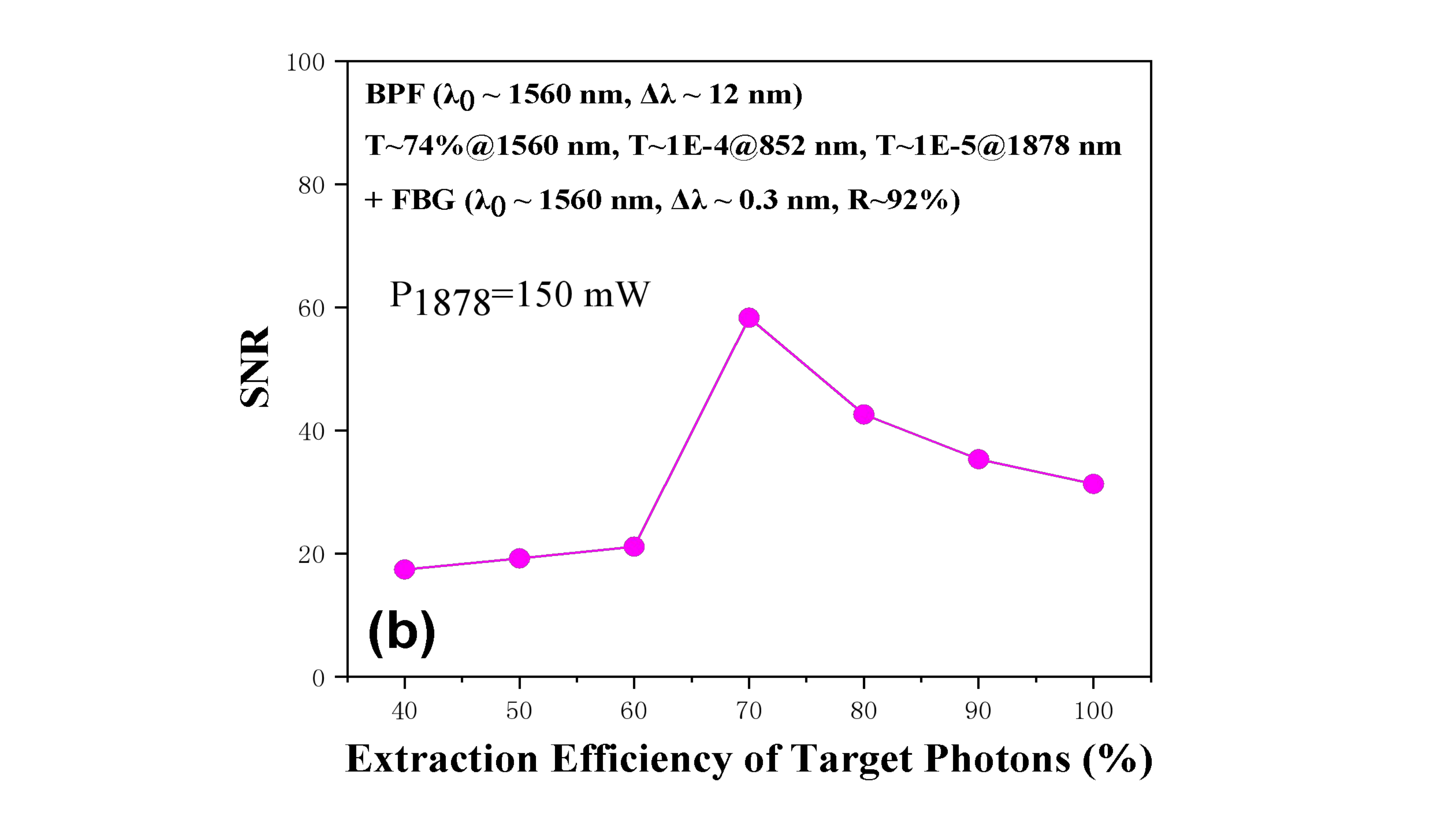}
} \vspace{-0.1in} 

\end{figure}

\begin{figure}[h]
\setlength{\belowcaptionskip}{-0.5cm}
\vspace{-0.1 in}
\centerline{
\includegraphics[width=95mm]{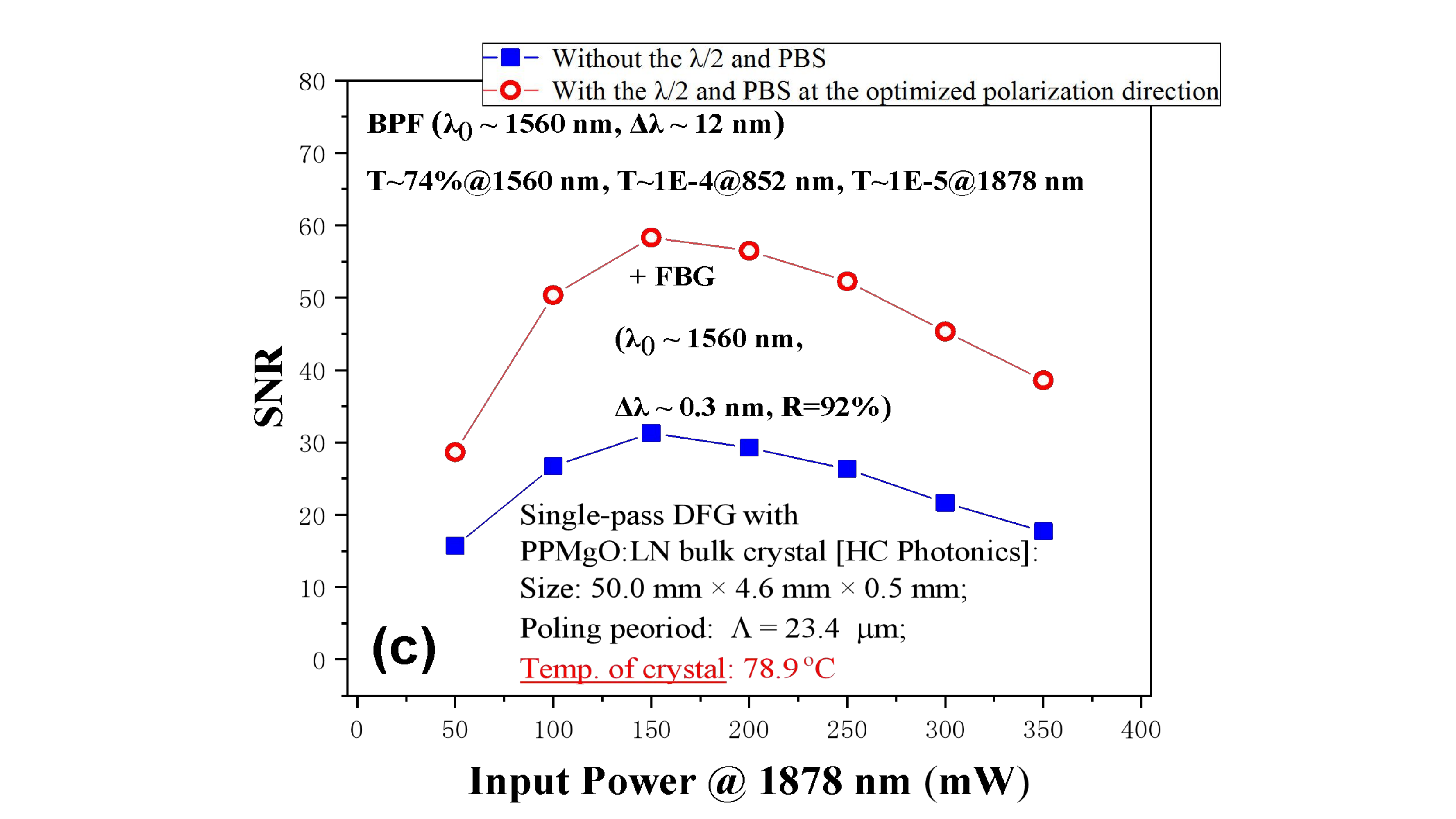}
} \vspace{0in} 

\end{figure}

\begin{figure}[h]
\setlength{\belowcaptionskip}{-0.5cm}
\vspace{0in}
\centerline{
\includegraphics[width=95mm]{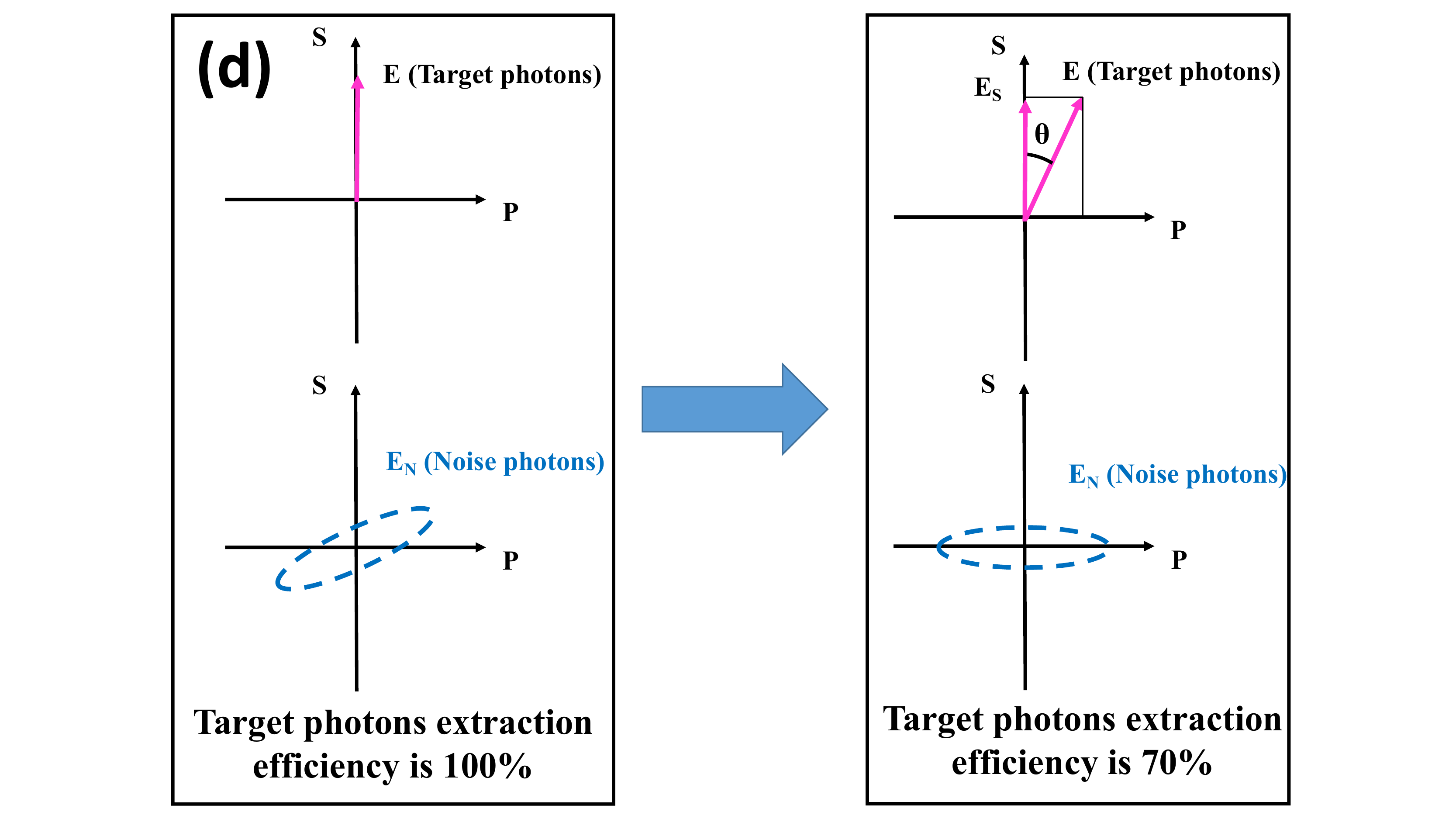}
} \vspace{0.1in} 
\caption{Improving SNR by changing polarization of output photons. (a) The experimental setup for polarization optimization. BPF: a 1560-nm band-pass filter with  $ \sim $ 12-nm FWHM bandwidth (T $ \sim $ 74\%@1560 nm, T  $ \sim $  1E-4@852 nm, T  $ \sim $  1E-5@1878 nm); Circ.: single-mode fiber circulator; FBG: 1560-nm fiber Bragg grating filter with 0.3-nm FWHM bandwidth (R $ \sim $ 92\%@1560 nm); ${{{\lambda}/2}}$: half-wave plate; PBS: polarization beam splitter cube; SPCM: single-photon counting module. (b) SNR versus target photons' extraction efficiency. When the pump laser is 150 mW, the SNR can be improved to 58.3 with  $ \sim $  70\% of extraction efficiency. (c) SNR versus 1878-nm pump power. The blue solid squares represent SNR without the ${{{\lambda}/2}}$ plate and PBS, and the red open circles represent SNR with the ${{{\lambda}/2}}$ and PBS at the optimized polarization direction (extraction efficiency is  $ \sim $  70\%). The SNR increases from 31.3 to 58.3. (d) The polarization of target photons and noise photons versus different extraction efficiency. E is linear polarization for the target photons and EN is elliptical polarization for noise photons (anti-Stokes photons). The left side shows the polarization of target photons and noise photons when the extraction efficiency is 100\%, and the right side shows the polarization when the extraction efficiency is  $ \sim $  70\% (after polarization rotation with an angle $\theta$ $ \sim $ 33 $^{\circ}$).}
\label{Fig 6}
\vspace{0.15in}
\end{figure}

 \noindent SNR increases from 31.3 to 58.3. Fig. 6 (d) shows  polarization of target photons and anti-Stokes photons before  and after polarization optimization. The left side shows the polarization of target photons and noise photons when the target photons' extraction efficiency is  $ \sim $ 100\%, while the right side shows the polarization when the target photons' extraction efficiency is  $ \sim $ 70\%. The polarization rotation angle  $\theta$ $ \sim $ 33$^{\circ}$. The picture in Fig. 6(d) can correctly describe the measured data in Fig. 6(b).

\section{SFG from 1560-nm to 852-nm photons: Experiment and results}

In this part, we convert attenuated coherent state from 1560 nm to 852 nm by SFG in PPMgO:LN crystaland analyze the noise photons induced by the strong 1878-nm pump laser beam. Fig. 7 (a) is diagram of single-photon-level SFG photonic conversion. The experimental setup for SFG is shown in Fig. 7 (b). To simulate single-photon pulses, a compact DFB diode laser at 1560 nm, an AOM, and a strong attenuator are employed to chop and weaken 1560-nm continuous-wave laser beam to 1-MHz repetition rate and 500-ns duration square-wave periodical


\begin{figure}[h]
\setlength{\belowcaptionskip}{-0.5cm}
\vspace{-0.1 in}
\centerline{
\includegraphics[width=75mm]{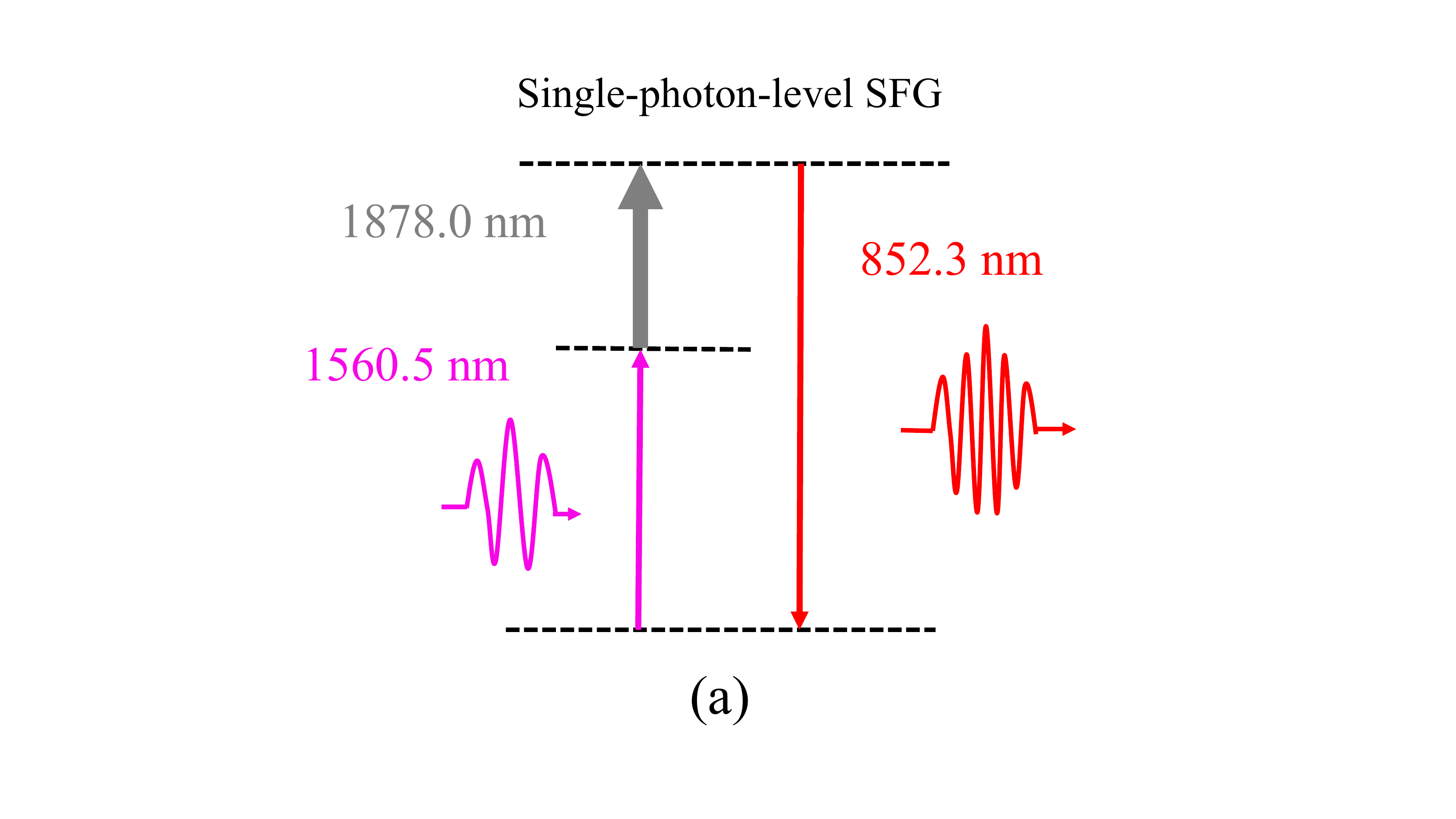}
} \vspace{-0.3in} 

\vspace{0in}
\end{figure}

\begin{figure}[h]
\setlength{\belowcaptionskip}{-0.5cm}
\vspace{0 in}
\centerline{
\includegraphics[width=95mm]{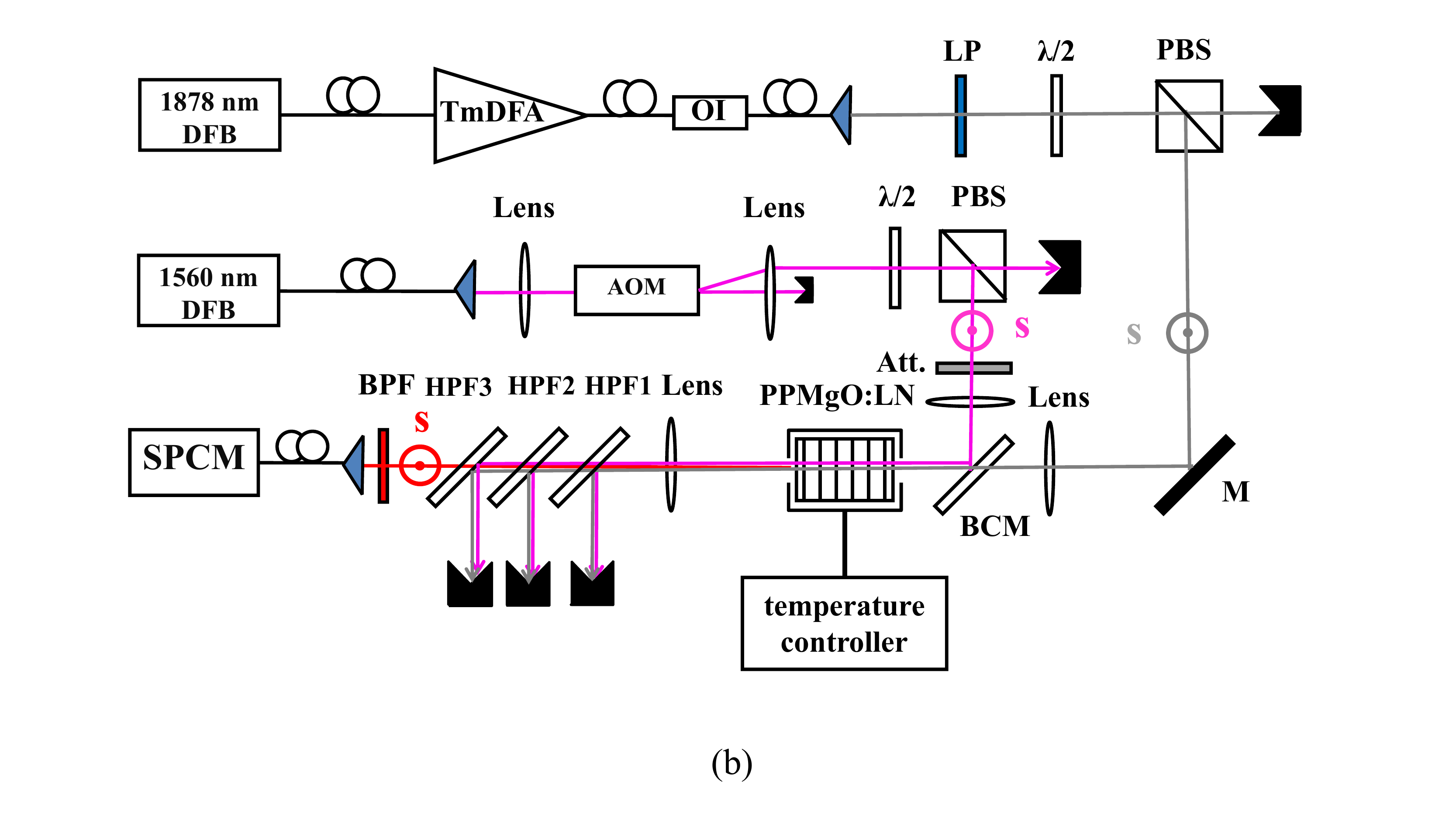}
} \vspace{0in} 
\caption{(a) Diagram of single-photon-level SFG photonic conversion. (b) The experimental setup for DFG, the s-polarized 1560-nm photons are converted to 852-nm photons with PPMgO:LN crystal by SFG pumped by the s-polarized 1878-nm strong laser beam. DFB: distributed feedback diode laser; TmDFA: Thulium-doped fiber amplifier; OI: optical isolator; AOM: acousto-optic modulator; ${{{\lambda}/2}}$: half-wave plate; PBS: polarization beam splitter cube; Att.: attenuator; s: s polarization; BCM: beam-combination mirror; HPFs: high-pass filters (852-nm high-transmission and 1560-nm and 1878-nm high-reflection mirrors); BPF: 852-nm band-pass filter (transmittances: $ \sim $75\%@852 nm, $ \sim $44\%@1560 nm and $ \sim $35\%@1878 nm, FWHM bandwidth: $ \sim $0.5 nm and OD=4 for 853 nm $ \sim $ 1060 nm and 600 nm $ \sim $ 851 nm); SPCM: single-photon counting module; LP: low pass filter.}
\label{Fig 2}
\vspace{0.1in}
\end{figure}

 \noindent   optical pulses with a mean photon number per pulse is $ \sim $ 1. The strong pump field is same as DFG experiment. Then, the same 50-mm-long PPMgO:LN crystal is used as SFG crystal. We choose two f = 75 mm lenses to focus 1878-nm and 1560-nm laser beams separately, then combine two focused beams in the PPMgO:LN bulk crystal, to avoid the chromatic aberration in the case of using one focusing lens for these two wavelengths. The waist spots radii of 1560 nm and 1878 nm are $ \sim $ 41.4 ${{\mu}}$m and $ \sim $ 48.3 ${{\mu}}$m, respectively. According to the B-K focus factor, the optimal waist spots radii for 50-mm-long crystal of 1560 nm and 1878 nm are $ \sim $ 45.1 ${{\mu}}$m and $ \sim $ 49.8 ${{\mu}}$m respectively. After crystal, a 75-mm lens is used to collimate the output laser, and output lasers are spectrally separated by high-pass filters (HPFs) (852-nm high-transmission and 1560-nm and 1878-nm high-reflection mirrors). Then the 852 nm photons passed through the 852-nm BPF (please see the caption of Fig.7). Finally, the 852 nm photons are counted by a SPCM (PerkinElmer Optoelectronics; SPCM-AQR-15; quantum efficiency  $ \sim $ 48\% and dark count rate $<$ 50 counts/s).

In the single-pass SFG experiment, we find the optimized phase matching by adjusting the temperature of PPMgO:LN crystal. When 1878-nm pump power is $ \sim $ 450 mW, the quasi-phase matching temperature is $ \sim $ 77.6$^{\circ}$C. The normalized results are shown in Fig. 8.

\begin{figure}[h]
\setlength{\belowcaptionskip}{-0.5cm}
\vspace{-0.05 in}
\centerline{
\includegraphics[width=95mm]{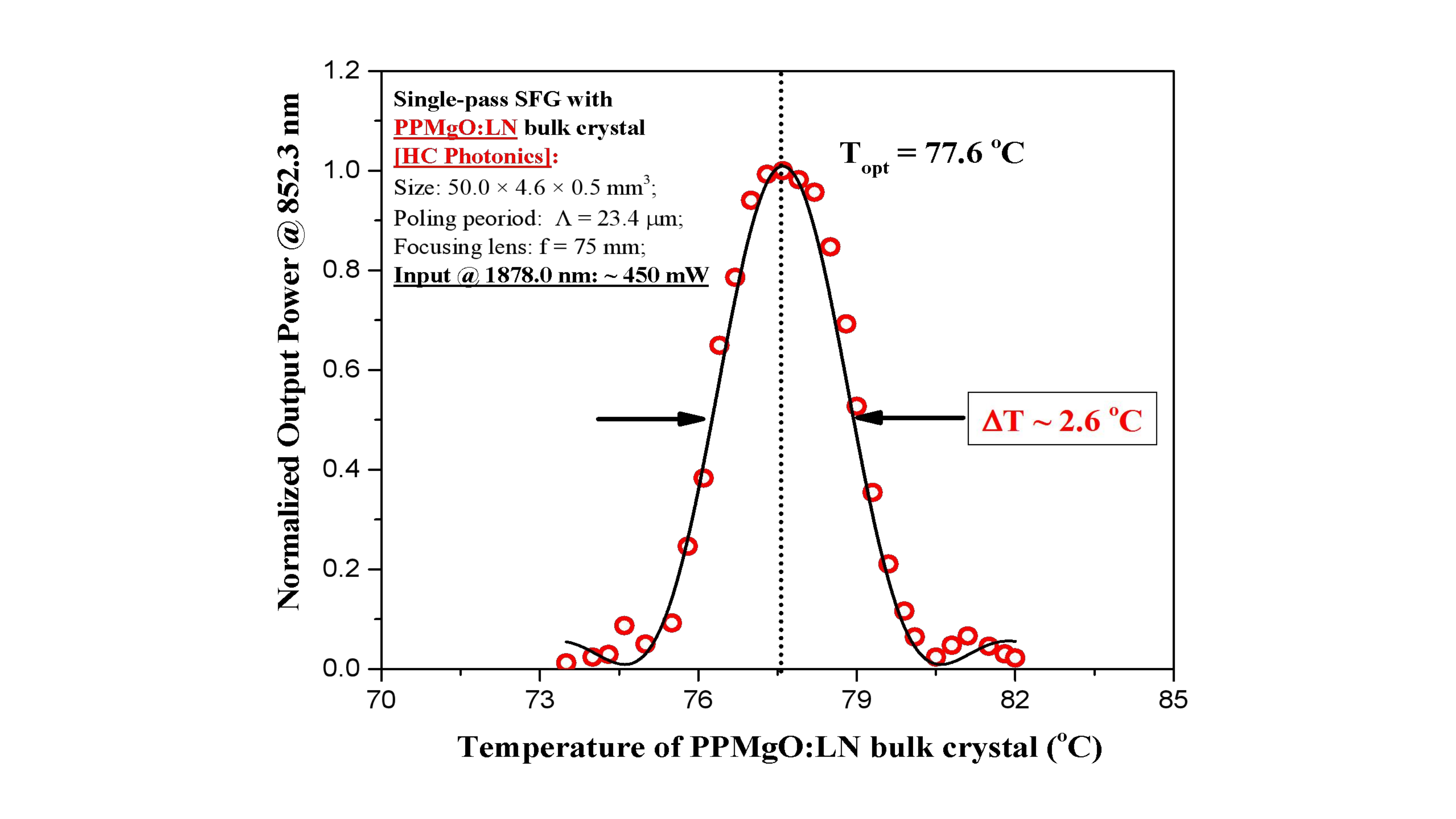}
} \vspace{-0.05in} 
\caption{Dependence of 852.3-nm output power upon the crystal's temperature in the case of SFG. The open circles represent the experimental data, while the solid line is theoretical fitting curve. The optimal quasi-phase-matching temperature is $ \sim $ 77.6$^{\circ}$C, and the temperature bandwidth is $ \sim $ 2.6$^{\circ}$C.}
\label{Fig 2}
\vspace{0.3in}
\end{figure}

For single-photon-level frequency conversion it is known that noise (unwanted photons) at the target wavelength can be generated by the strong pump field. Many groups have also characterized the noise in the SFG experiments, for example, ref.[17]. Fig.9 shows the overview of SFG photonic conversion process from $ {\lambda}_{2}$ = 1560.5 nm to $ {\lambda}_{1}$ = 852.3 nm with the $ {\lambda}_{p}$ = 1878.0 nm pump laser, the strong pump induced nonlinear processes (SPDC and SRS), and the cascaded conversion. First, the pump laser will induce the SPDC and SRS as well as second-harmonic generation (SHG) processes in the crystal. The SPDC photons locate at the longer wavelength (marked by the purple part). The SRS photons distribute on both sides of the pump light (marked by dark grey part). The SPDC, SRS and SHG photons, as well as the noise photons (generated by the cascaded SFG process and marked by the green part) located at the target output can be removed by using the HPFs and the BPF.

\begin{figure}[h]
\vspace{-0.3in}
\centerline{
\includegraphics[width=95mm]{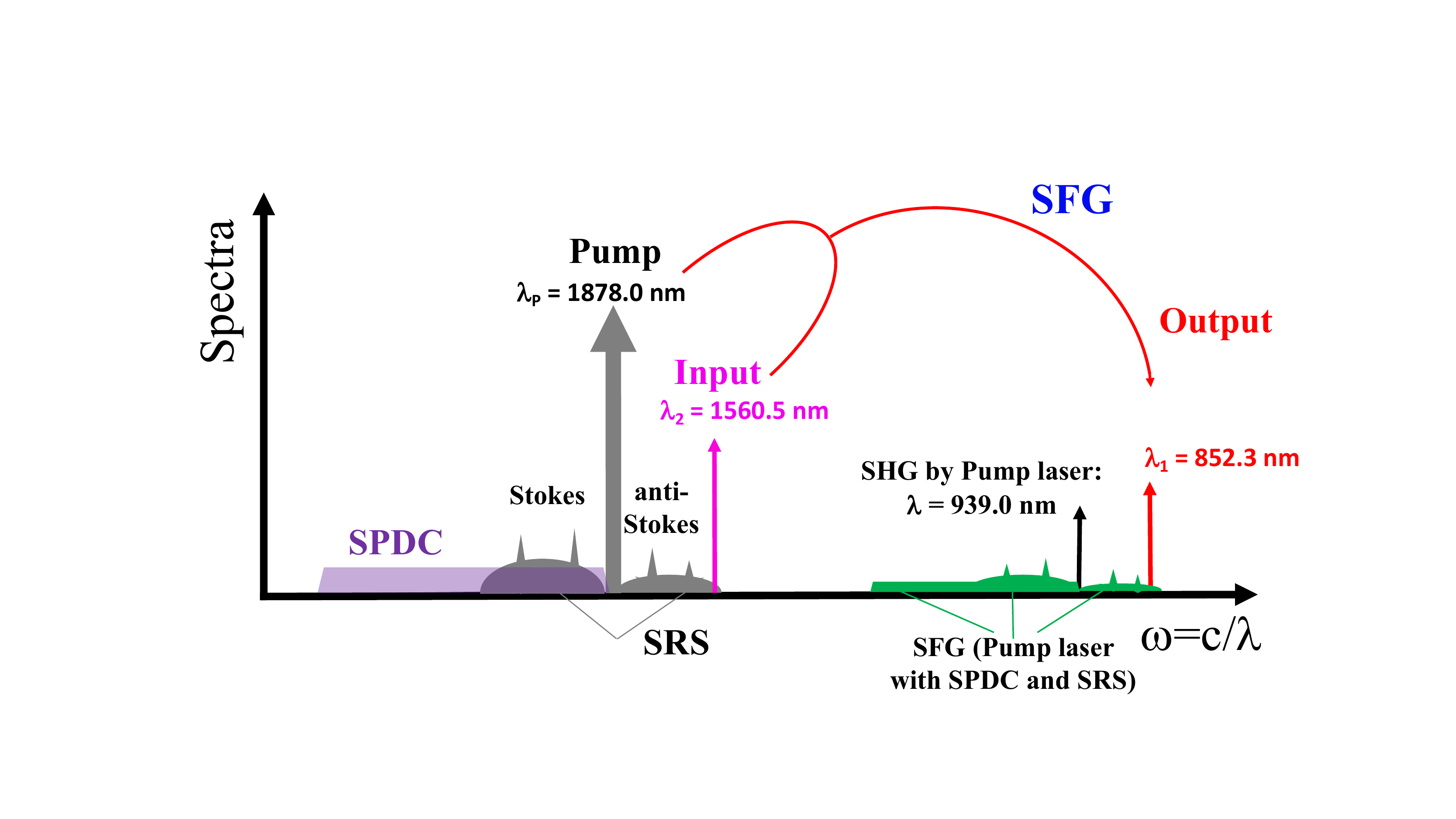}
} \vspace{-0.2in} 
\caption{Overview of SFG photonic conversion process from 1560.5 nm to 852.3 nm with the 1878.0-nm pump laser, the strong pump induced nonlinear processes (SPDC and SRS), and the cascaded conversion. SPDC: the spontaneous parametric down-conversion (marked by the purple part); SRS: the spontaneous Raman scattering (Stokes and anti-Stokes photons, and marked by dark grey part); SHG: second-harmonic generation (939.0 nm and marked by the black arrow). The SPDC, SRS and SHG photons, as well as the noise photons (generated by the cascaded SFG process and marked by the green part) locate at the target output can be removed by using the HPFs and the BPF.}
\label{Fig 4}
\vspace{-0.2in}
\end{figure}

Different SNR and conversion efficiency can be obtained by changing pump power. The definition of efficiency is the same as that of the DFG. These results are shown in Fig. 10. The blue solid squares represent SNR, while the red open circles represent the SFG conversion efficiency. The solid lines represent the fittings to Eq. (1). The maximum SFG internal conversion efficiency is $ \sim $ 1.9\%. The maximum SNR is 38.8 with 150 mW of pump laser beam. If we use the filters with narrower FWHM, the SNR will be significantly improved. And we will also further improve the SNR by rotating polarization of output photons like DFG experiment. As the same as DFG results, the reason for efficiency saturation is caused by the consumption ratio of signal and pump photons, and the mode matching of input photons (signal photons and pumping photons) is the main factor for the lower conversion efficiency. In addition, the inverse DFG (from 852-nm to 1560-nm photons) is inevitable in the SFG experiment. This means that some of the 852-nm photons generated by the SFG may be further converted back to 1560-nm photons. Although the conversion efficiency of the inverse DFG is much less than the target SFG, that factor may limits our SFG efficiency.

\begin{figure}[h]
\vspace{-0.05in}
\setlength{\belowcaptionskip}{-0.5cm}
\centerline{
\includegraphics[width=95mm]{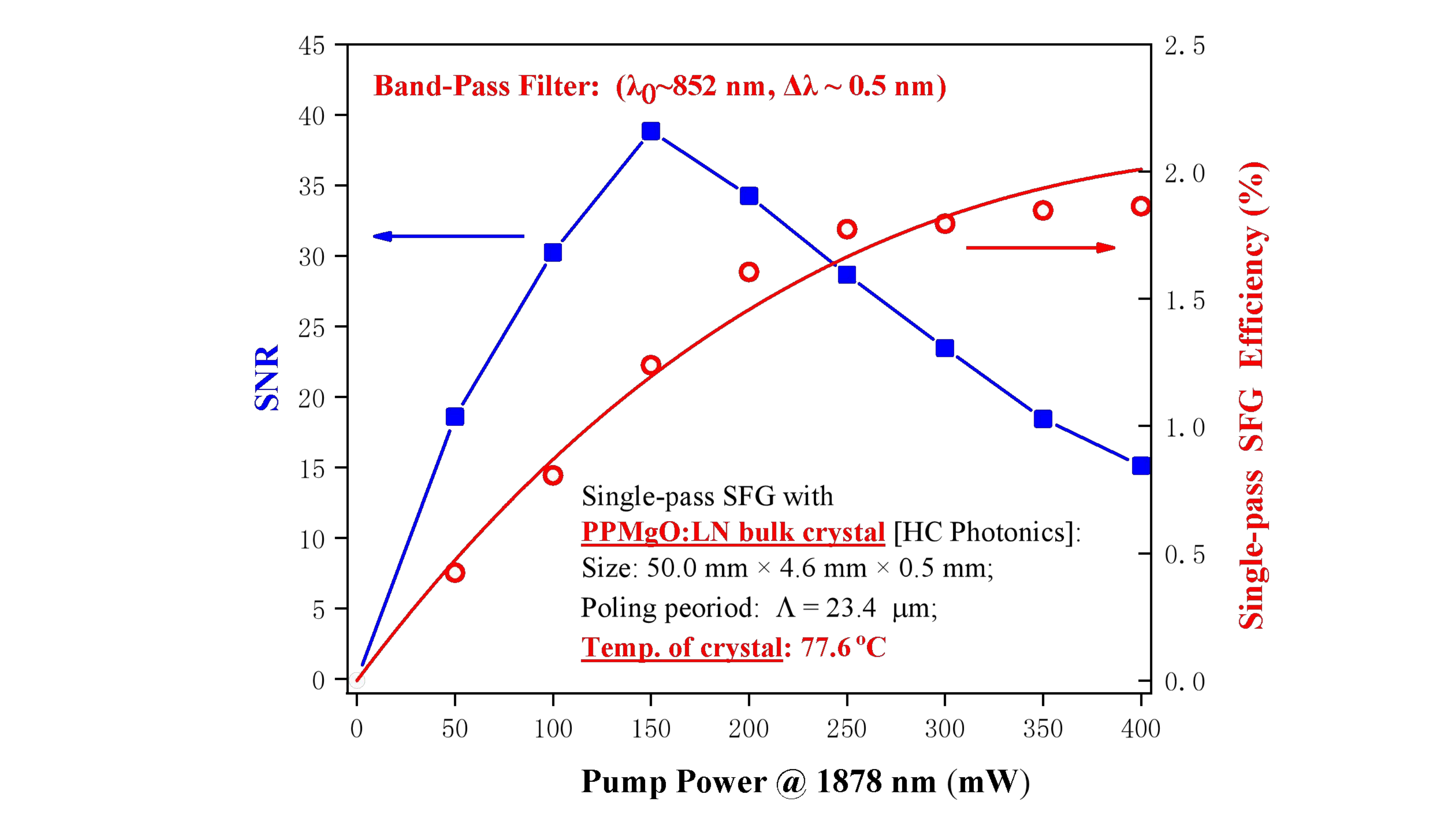}
} \vspace{0.1in} 

\caption{Experimental data for SNR and SFG conversion efficiency versus the 1878-nm pump power. The blue solid squares represent SNR, while the red open circles represent SFG efficiency.}
\label{Fig 11}
\vspace{0.2in}
\end{figure}

\section{Conclusions}

We report the two-way single-photon-level frequency conversion between 852 nm and 1560 nm. First, the 852-nm photons at cesium D2 line have been converted (DFG) to 1560-nm photons at telecom C-band in a PPMgO:LN bulk crystal with strong pump field at 1878 nm. The DFG internal conversion efficiency is $ \sim $ 1.7\% with 400-mW pump laser mixing attenuated 852-nm laser pulses with a mean photon number per pulse is $ \sim $ 1, and the maximal SNR is 58.3. We can improve SNR by using narrower bandwidth filters and changing the polarization of the noises. Finally, SNR can be nearly doubled than before. Then the 1560-nm photons at telecom C-band have been converted (SFG) to 852-nm photons at cesium D2 line with the same conditions. The SFG internal conversion efficiency is $ \sim $ 1.9\% with 400-mW pump laser mixing attenuated 1560-nm laser pulses with a mean photon number per pulse is $ \sim $ 1, and the maximal SNR is 38.8. In addition, we analyze the noise characteristics in conversion processes. The low conversion efficiency is mainly due to the mismatch of the waist spots and the mode volumes of pump laser and the signal photons. In addition, if we use waveguide in this experiment, the efficiency will be greatly improved. This is benefited by the special structure of the waveguide.

The two-way conversion based on the same crystal can build a bridge between the quantum systems and realize the information transmission. This scheme can be extended to real single photons case, and combining our single photon source to realize the quantum network for ultra-low loss transmission in the future.

\vspace{0.25in}

 \quad\qquad \quad\textbf{ACKNOWLEDGMENTS}

\vspace{0.15in}
\vspace{0.05in}

This work is supported by National Natural Science Foundation of China (11774210, 11974226 and 61875111); National Key R\&D Program of China (2017YFA0304502); Shanxi Provincial Graduate Innovation Project (PhD Candidates) (2019BY016) and Shanxi Provincial 1331 Project for Key Subject Construction.

\end{document}